\documentclass[structabstract]{aa}  
\usepackage{graphicx}
\usepackage{txfonts}
\usepackage{natbib}
\bibpunct{(}{)}{;}{a}{}{,}

\begin{document}
   \title{SCG0018-4854: a young and dynamic compact group \thanks{Based on observations collected at the European Southern
Observatory, Cerro Paranal, Chile (proposal: 267.A-5722(A))}}

   \subtitle{I. Kinematical analysis}

   \author{
	V. Presotto
          \inst{1,2}
          \and
          A. Iovino\inst{2}
	  \and
	  E. Pompei\inst{3}
          \and
	  S. Temporin\inst{4}
          }

   \institute{
		Universit\'a degli Studi dell'Insubria,
              Via Valleggio 11, 22100 Como, Italy\\
              \email{valentina.presotto@brera.inaf.it}
         \and
             INAF - Osservatorio Astronomico di Brera
	     Via Brera 28, 20122 Milan, Italy
         \and
             European Southern Observatory (ESO)
	     Alonso de Cordova 3107, Santiago, Chile
         \and
             Institut f\"ur\ Astro- und Teilchenphysik, Universit\"at\ Innsbruck,
	     Technikerstrasse 25, 6020 Innsbruck, Austria
             }

   \date{Received Month Day, Year; accepted Month Day, Year}

 
  \abstract
   {}
   {It is widely recognized that processes taking place inside group environment could be among the main 
   drivers of galaxy evolution. Compact groups of galaxies are in particular good laboratories for studying 
   galaxy interactions and their effects on the evolution of galaxies due to their high density and low 
   velocity dispersion. SCG0018-4854 is a remarkably high galaxy density and low velocity dispersion 
   group with evidence of a recent interaction. It has been detected and analyzed at different wavelengths, 
   but its kinematics has not yet been studied in detail.}
   {We obtained VLT FORS2 optical observations and we present spectroscopic and photometric evidence of how 
   dramatically galaxy interactions have affected each of the four member galaxies.}
   {We found peculiar kinematics for each galaxy and evidence of recent star formation. In particular, 
   the gas and stellar radial velocity curves of two galaxies are irregular with a level of asymmetry 
   similar to that of other interacting galaxies. We discovered the presence of a bar for NGC 92 therefore 
   revising a previous morphological classification and we obtained spectroscopic confirmation of a 
   galactic-scale outflow of NGC 89.}
   {Peculiar kinematics and dynamic consideration lead to a rough estimate of the age of the latest
   interaction: $\tau \approx 0.2-0.7 \, \mathrm{Gyr}$, suggesting that SCG0018-4854 is a young and 
   dynamical group.}


   \maketitle
%

\section{Introduction}

Compact groups (CGs) are among of the smallest and densest systems of galaxies of the universe, 
furthermore they are characterized by low velocity dispersions of the order of 
$\approx 200\, \mathrm{km \, s^{-1}}$ \citep{1992ApJ...399..353H}. This is why they are excellent 
laboratories for studying galaxy interactions and their effects on the evolution of galaxies. 
\citet{1991ApJS...76..153R} showed that 30\% of spiral galaxies in CGs has distorted rotation 
curves and another 30\% has irregular velocity patterns indicating that tidal interactions are frequent 
and persisting in CGs. Moreover, \citet{1994ApJ...427..684M} demonstrated that 43\% of the galaxies in 
their compact group sample shows morphological and/or kinematical distortions indicative of interactions 
and/or mergers. Since these pioneering studies it has been clear that interactions play an important 
role in the evolution of groups. According to current dynamical models, galaxies belonging to groups 
should interact violently and merge on a relatively short time scale 
\citep{1987ApJ...322..605S,1985MNRAS.215..517B,1996ApJ...458...18G}. Yet, strong mergers in CGs are 
rare \citep{1994ApJ...427..684M} and a lot of effort has been made to establish the origin and fate of 
such systems. Recent studies \citep{2007AJ....133.2630C} have shown that galaxies in compact groups are 
more likely to merge under dry conditions, after they have lost most of their gas in interactions among 
galaxies. These hypothesis is corroborated by HI studies; \citet{2001A&A...377..812V} have found out 
perturbed distributions of neutral hydrogen in CGs. This results indicate that the evolution of CGs is 
mainly dominated by galaxy interactions, continuous tidal stripping and/or gas heating.

It is therefore interesting to reconstruct the evolution and the interaction history of a CG throughout 
the properties of its galaxy members and compare them with group properties as a whole. Kinematical 
and dynamical information are fundamental to separate between different interaction histories. 
\citet{1998ApJ...507..691M} defined some kinematic indicators to distinguish between merging and 
interaction, such as misalignment between the kinematic and photometric axis of gas and stellar components, 
double gas components, warping and other peculiarities. However, there are still open questions about how 
many interactions take place and how strong and frequent they are. 

In this work we focus on SCG0018-4854, a spiral-only compact group characterized by a really dense 
environment in which member galaxies have clearly interacted.

The plan for the article is the following. In Sect.~\ref{sec:SCG0018-4854}, we present the 
characteristics of SCG0018-4854. In Sect.~\ref{sec:obs}, we describe our observations and 
the reduction process. In Sect.~\ref{sec:method}, we explain the methods used for our analysis. 
The results are presented in Sect.~\ref{sec:results}, followed, in 
Sect.~\ref{sec:Discussion}, by a short discussion and our conclusions in 
Sect.~\ref{sec:Conclusions}.


\section{SCG0018-4854}
\label{sec:SCG0018-4854}

SCG0018-4854 is an extremely compact group of four late-type galaxies: NGC 92,
NGC 89, NGC 87 and NGC 88. It was already listed in Rose's catalogue
\citep{1977ApJ...211..311R} and it is part of the Southern Compact Group
catalogue \citep{2000ASPC..209...25I}. It is located at 
$\mathit{v} \approx 3240 \, \mathrm{km \, s^{-1}}$ which means 
$\mathit{D} \approx 45\mathrm{h_{70}} \, \mathrm{Mpc}$, 
adopting $\mathit{h_{70}} = \mathrm{H_{0}}/70 \, \mathrm{km \, s^{-1} \, Mpc^{-1}}$. 
All galaxy members are within a projected distance of $\approx$ 1\farcm6
from the center of the group, where $1\arcsec$ corresponds to $225 \mathrm{pc}$. A possible fifth
member, ESO198-G013, is located at a distance of $\sim\, 15\arcmin$ in the E direction.
Furthermore, SCG0018-4854 has a remarkably low velocity dispersion, $\approx 120
\, \mathrm{km \, s^{-1}}$ and high density, $4.88 \, \mathrm{gal \, Mpc^{-3}}$ in log scale. Diffuse HI and X-ray emission has recently been detected confirming its bound nature
\citep[see][]{2007A&A...473..399P,2008A&A...484..195T}. The former data reveal a
common HI envelope and a possible bridge between two galaxies, NGC 92 and NGC
88; according to \citet{2001A&A...377..812V} this HI distribution is typical of
groups in phase 3b, i.e. with HI stripped from the galaxies and distributed in the
common potential or in tidal tails. X-ray observation have revealed the presence of a hot
intergalactic medium, $\mathit{kT} \approx 0.2 \, \mathrm{keV}$, in which the galaxies 
are embedded. The
presence of a warm and hot intergalactic diffuse medium is important not only as
an evidence of the gravitationally bound nature for SCG0018-4854, but also a
proof of the galaxies interaction, even though it alone cannot explain tidal structures. 
Actually, all four close members show a disturbed morphology and a late-type appearance with 
intense activity, as testified by early spectroscopic data \citep{2000AJ....120...47C}. 
Here below we list the main properties of each galaxy while other information can be found in 
Table~\ref{tab:SCG0018}:

\begin{itemize}
 \item NGC 92 is a SAa pec starburst, also classified as a LINER by
 \citet{2000AJ....120...47C}. It shows a ring of star-forming regions in its central
 part and it has an extended tidal tail in SE direction with many 
 HII regions \citep{2005ASSL..329P..78T}. A peculiar extension in the tail direction is also detected 
 in X-ray \citep{2008A&A...484..195T}. Moreover, it has been suggested to contain a double 
 nucleus \citep{1981A&A....98..223D}. 
 The galaxy has clearly undergone an interaction with one or several of the group members.
\item NGC 89 is a SB0/a pec and it has a Sy2 nucleus and  
H$\alpha$-emitting extra-planar features with a filamentary structure on both sides
of the disk, including a $\sim$ 4 kpc-long filament with a jet-like structure extending in NE direction 
\citep{2005ASSL..329P..78T}. It has also a ring of star forming regions around the nucleus. 
Interestingly, NGC 89 is the only galaxy of the group that was not detected in 
HI \citep{2007A&A...473..399P}. Probably it has lost its neutral gas after interacting with 
other group members and the IGM.
\item NGC 88 is the third spiral, SBab pec, of the group with a LINER nucleus. It forms an HI bridge 
with NGC 92 \citep{2007A&A...473..399P}
\item NGC 87 is a nearly face-on irregular galaxy, based on its axis ratio. 
It has many distinct regions of strong star formation.
 \end{itemize}

\begin{table*}
\caption{SCG0018-4854 galaxy properties. RA and DEC data are taken from Nasa/IPAC Extragalactic 
Database, while morphology \& activity are taken from \citet{2007A&A...473..399P}. 
References: (1) \citealt{1981A&A....98..223D}; (2) \citealt{2005ASSL..329P..78T} and (3) 
\citealt{2007A&A...473..399P}.}
\label{tab:SCG0018}
\centering
\begin{tabular}{l c c c c c c}
\hline\hline
GALAXY & RA & DEC & V  & Morphology & Activity & Notes\\
       & (J2000.0) & (J2000.0) & ($\mathrm{km \, s^{-1}}$) &  &  & \\
\hline
    NGC 92 & 00:21:32 & -48:37:29 & 3378$\pm$10 &   SAa pec & SBG & tidal tail \\
           &          &           &      &       &       & double nucleus? $^{1}$\\   
    NGC 89 & 00:21:24 & -48:39:55  & 3287$\pm$10  &  SB0/a pec & Sy2 & H$\alpha$ galactic-scale
    extra-planar features $^{2}$ \\
           &          &           &      &       &       & no HI detection $^{3}$\\   
    NGC 87 & 00:21:14  & -48:37:42  & 3415$\pm$23  &  Irr & SFG & -- \\
    NGC 88 & 00:21:22  &  -48:38:25 & 3439$\pm$19  &  SBab pec & LINER & -- \\
\hline
\end{tabular}
\end{table*}

All these properties make SCG0018-4854 one of the few spiral-only and most active groups among all 
compact groups. It can be considered as the southern counterpart of HCG16, the most active group of 
all Hickson Compact Groups. This group is located at a similar redshift and it is composed of four 
peculiar late-type galaxies, as described by \citet{1982ApJ...255..382H}, recently other three members 
have been added \citep{1996ApJ...463L...5R}. It has been detected both in HI and X-ray revealing the 
presence of a common medium \citep{2001A&A...377..812V,2003A&A...398....1B}. All galaxy members show 
signs of interaction, such as tails and double nuclei, and different activities ranging from starburst 
to AGNs \citep{1999AJ....117.1657D}. The interaction affected also the velocity field of each galaxy, 
as shown by \citet{1998ApJ...507..691M} who observed the presence of peculiar kinematics.

SCG0018-4854 has been detected and analyzed at different wavelengths, but its kinematics 
has not yet been studied in detail. In this paper we present first results of the kinematical analysis 
of this group and we compare its properties with those of HCG16. 
In a forthcoming paper (Temporin et al. in preparation) we will analyze in detail the 
photometric and morphological properties of each galaxy of SCG0018-4854, focusing on the likely age of
the stellar populations and their correlation with morphological structures.

\section{Observation and data reduction}
\label{sec:obs}

\begin{figure}
 \centering
\resizebox{\hsize}{!}{\includegraphics{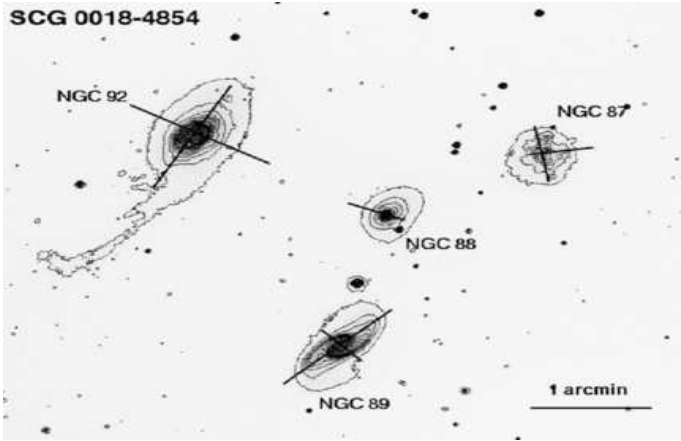}}
 \caption{$R$-band image of SCG0018-4854 with the slits and isointensity contours overlayed. 
 North is up and east to the left.}
 \label{fig:slit_pos}
\end{figure}

The spectroscopic data have been obtained with the FORS2 spectrograph at the VLT-UT1, 
\citep{1998Msngr..94....1A}, during an observing run in September 2001. The detector 
was a $2048 \times 2048 \, \mathrm{px}$ CCD with a
scale of $0.201 \arcsec \, \mathrm{px^{-1}}$ (we used the standard resolution collimator).
We acquired spectra using the Spectroscopic Mask Mode (MXU). This kind of set-up
offers the possibility to use a mask with slits of different length, width,
shape and inclination suiting the observation demands. We used two different masks,
one for all spectra along the major axis and the other for all  
spectra along the minor
axis. We couldn't observe the major axis of NGC 88 due to overlaps of the slits
in the same mask. We needed two different slits to cover the entire galaxy
extent of NGC 92 (both axis) and NGC 89 (major axis). The data consist of four
exposures of $2699 \, \mathrm{s}$ for both masks, covering the major and the 
minor axis. All slits are
$1 \arcsec$-wide and they have different length, position and inclination with
respect to the dispersion axis. We used the GRISM-1400V with a dispersion of 20.8 \AA\ mm$^{-1}$, 
corresponding to 0.49 \AA\  px$^{-1}$. The grism peak is at
5200 \AA\ and the spectral coverage of each object spectra is $\approx 1000$
\AA\ in the 4500-5900 \AA\ wavelength range according to the slit position
with respect to the center of the CCD. Figure~\ref{fig:slit_pos} and 
Tables~\ref{table:slitM}, \ref{table:slitm} show the slit positions and properties. A set of
He-Cd-Hg-Ne lamp spectra have been taken for each observing night. The average
seeing was about $0.7 \arcsec$ throughout our exposures. The individual frames were 
pre-reduced (bias subtraction, flat-field correction, etc.) using standard 
IRAF \footnote{IRAF is distributed by the National Optical Astronomy Observatories, 
which are operated by the Association of Universities for Research in Astronomy, Inc., 
under cooperative agreement with the National Science Foundation.} image processing packages. 
The wavelength calibration was made using the IRAF TWODSPEC.LONGSLIT package. Our spectra
have different inclination angles with respect to the dispersion axis. In order to rectify
them we adapted the parameters of the task REIDENTIFY forcing it to identify and calibrate
each lamp emission-line row by row. We verified our final wavelength solution using
available skylines row by row. We adopt the rms of the $\lambda 5577$ \AA\ skyline
centroid distribution, $\approx 0.1$ \AA, as the calibration accuracy.  
The wavelength coverage for NGC 87 does not include any strong skyline so we verified
the calibration directly on lamp spectra. We obtained the same accuracy as above. We
found a systematic offset of $\approx 0.5$ \AA\ in the calibration of all spectra
along the major axis, probably due to instrument flexures \citep[see][]{2001AJ....121.2572G}.
The final spectral resolution turned out to be 2.25 \AA\ (FWHM of $\lambda 5577$
skyline), equivalent to a velocity of $\sigma\approx 51 \, \mathrm{km \, s^{-1}}$. 
The sky subtraction was
performed using a narrow region, $\approx 10 \, \mathrm{px}$, at both edges of the slit where there
was minimum galaxy contamination. Unfortunately, we could not remove the sky from the
spectrum along the major axis of NGC 87 because the galaxy covers the whole slit (a
longer slit would have caused problems of overlap with the others). Our attempts to
reproduce a good sky level failed, so we preferred not to subtract it. We extracted 1D
spectra binning each 2D spectrum along the spatial direction in order to keep a
signal-to-noise ratio (S/N) $\gid$ 6 at all radii. It should be noticed that in the
central pixels we have a $\mathrm{S/N} \gid 100$, having such high S/N data we just take into account the 
seeing when binning spectra in order to obtain uncorrelated spatial measure.

\begin{table}
\caption{Slit properties of the mask along the major axis. i=inclination angle with respect to 
the dispersion axis, PA=position angle with respect to the N direction.}
\label{table:slitM}
\centering
\begin{tabular}{cccccc}
\hline\hline
Gal slit & Width & Length & i & PA & $\lambda$ coverage \\
     & ($\arcsec$)  &  ($\arcsec$)  & (deg) & (deg) & (\AA~) \\
\hline
   NGC 92 & 1 & 2x39.8 & 0 & -29 &  4845-5895 \\
   NGC 89 & 1 & 2x35 & -18 & -47 &  4530-5585 \\
   NGC 87 & 1 & 30.5 & 39 & 10 &  4450-5490 \\
\hline
\end{tabular}
\end{table}

\begin{table}
\caption{Slit properties of the mask along the minor axis. i=inclination angle with respect to 
the dispersion axis, PA=position angle with respect to the N direction.}
\label{table:slitm}
\centering
\begin{tabular}{cccccc}
\hline\hline
Slit & Width & Length & i & PA & $\lambda$ coverage \\
     & ($\arcsec$)  &  ($\arcsec$)  & (deg) & (deg) & (\AA~) \\
\hline
   NGC 92 & 1 & 2x39.9 & 0 & 60 &  4620-5650 \\
   NGC 89 & 1 & 28 & -18 & 42 &  4850-5890 \\
   NGC 87 & 1 & 25 & 38.7 & -81 &  4450-5490 \\
   NGC 88 & 1 & 30 & 7.5 & 67 & 4620-5650 \\
\hline
\end{tabular}
\end{table}

\section{Analysis method}
\label{sec:method}

\subsection{Radial velocity curves}

Stellar kinematics was derived using the cross-correlation technique \citep{1979AJ.....84.1511T} 
with the IRAF RVSAO package. The cross-correlation was performed on a wavelength range of 
$\approx$ 900 \AA\ around the main absorption features of the Mg line triplet ($\lambda\lambda 5164, 
5173, 5184$ \AA). We decided to use the nuclear spectra of each galaxy as a template, thereby 
minimizing problems due to template mismatch. In this way we derived the relative velocities at 
various distances to the nucleus and so we constructed the stellar radial velocity curve. 
Error bars were derived by adding in quadrature the contributions from photon statistics, CCD read-out 
noise and calibration accuracy. It is worth noticing that stellar error bars are
always larger than gas ones because the cross correlation was performed in a
wavelength range with few absorption features. Furthermore those present in the spectra are not very intense, then the fit gives a large error.

In order to obtain the systemic velocity of each galaxy we performed a second cross-correlation with 
a radial velocity standard star extracted from the ESO archive. We used HD36003, a K star with an 
observing set-up that well matches that of our data. We assumed that the photometric center coincides 
with the dynamical one and we defined the systemic velocity of each galaxy to be that obtained with the 
cross-correlation with the standard star at the center of the galaxy along the major axis. All quoted 
velocities are corrected to the heliocentric reference system.

As already discussed, we could not measure the stellar kinematics of NGC 87 along the major axis 
because of the presence of the sky contamination. We decided to use the value of velocity at the 
center of the galaxy obtained from the minor axis as the systemic velocity.

The ionized gas kinematics was measured by a Gaussian fit to the main emission lines present in the 
spectra (H$\beta$, [OIII] $\lambda\lambda 4959, 5007$) using the IRAF ONEDSPEC.SPLOT task. We derived 
the errors adding in quadrature both the calibration accuracy and the formal errors given as output by the 
fitting routine. 

We discovered the presence of multiple components in the emission line profile of 
NGC 92 and NGC 89. In the case of NGC 92 we found two components and we used the deblending function 
of SPLOT to separate them, errorbars are the same as above. For NGC 89 the emission profile is quite 
complex, it could be the result of three or more overlapping components, see Sect. 
\ref{sec:kin89}. We used the task NGAUSSFIT in order to better deblend them. In this case 
the error bars for gas velocities are obtained as the sum in quadrature of four different sources: 
(1) the finite slit width introduces an uncertainty of 1.2 \AA. (2) the rms of the peak centroid among 
a sample of different fitting solutions obtained perturbing initial centroid guesses ranges between
0.01 and 0.5 \AA\ according to the component. (3) the error of the best fit ranges between 
0.001 and 0.5 \AA. (4) Finally the calibration accuracy is 0.1 \AA. All these terms add up to a 
velocity error that ranges between $65 \, \mathrm{and} \, 85 \, \mathrm{km \, s^{-1}}$ according to 
the component.

\subsection{Asymmetry Parameter}

We were interested in revealing any sign of interactions and how much they affected the kinematical 
properties of the galaxies. As we said, most galaxies in groups have irregular or anomalous rotation curves 
so we tried to quantify any asymmetry. As an indicator we used the asymmetry parameter, \textit{AP}, 
as described in \citep{2001AJ....121.1886D}:

\begin{equation}
\mathit{AP}=\sum\frac{\left|\left|\mathit{V(R)}\right|-\left|\mathit{V(-R)}\right|\right|}{\sqrt{\sigma^{2}(\mathit{R})+\sigma^{2}(\mathit{-R})}}\!\times\left[\frac{1}{2}\sum\frac{\left|\left|\mathit{V(R)}\right|+\left|\mathit{V(-R)}\right|\right|}{\sqrt{\sigma^{2}(\mathit{R})+\sigma^{2}(\mathit{-R})}}\right]^{-1}
\label{eq:AP} 
\end{equation}

This parameter takes into account how different is the behavior of the kinematically folded 
approaching and receding halves, so it quantifies possible asymmetries in the observed rotation curves. 
The error in \textit{AP} was estimated as follows: for each velocity measure, 
\textit{V(R$_i$)}, we generated a 
normal velocity distribution centred at \textit{V(R$_i$)} and having a dispersion $\sigma$ equal to the 
error-bar $\sigma$(R$_i$) of \textit{V(R$_i$)}; we drew a random value, \textit{V$_{new}$(R$_i$)}, 
out of it. Then we estimated the new \textit{AP} using \textit{V$_{new}$(R$_i$)} and $\sigma$(R$_i$). 
We repeated this procedure in a $\mathit{N} = 10000$ cycle and then we adopted as the \textit{AP} 
error the rms of the resulting \textit{AP} distribution, which resembles a normal one.

\section{Results}
\label{sec:results}
\subsection{Kinematics of NGC 92}\label{sec:kin92}

\begin{figure}
 \centering
\resizebox{\hsize}{!}{\includegraphics[angle=-90]{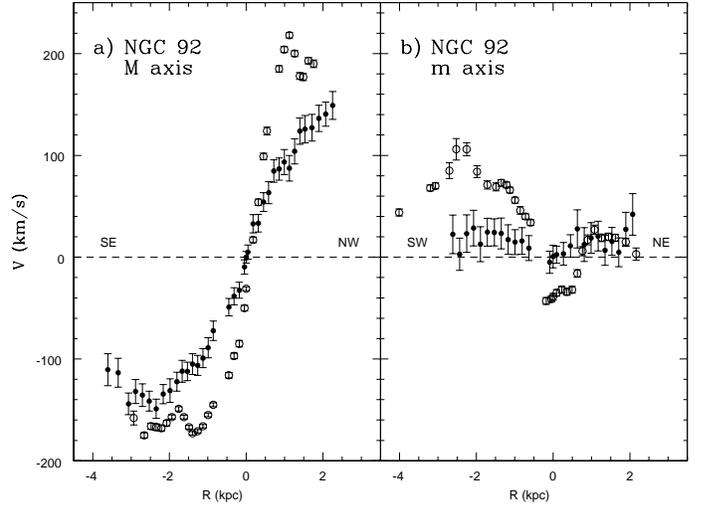}}
 \caption{Gas (open circles) and stellar (filled circles) kinematics along the Major (panel a) and 
 the minor (panel b) axis of NGC 92.}
 \label{fig:kin_A}
\end{figure}

In Fig.~\ref{fig:kin_A} we present the stellar (filled circles) and gas (open circles) radial 
velocity curves of NGC 92 along both the major (panel a) and the minor (panel b) 
axis. The kinematics along the major axis reveals two important features: i) at each radius the gas 
rotates faster than the stars. While the former reaches a maximal velocity 
$\mathit{V_{gas}} \approx 200-220 \, \mathrm{km \, s^{-1}}$, the latter reach a maximum at 
$\mathit{V_{*}} \approx 140\, \mathrm{km\, s^{-1}}$. ii) While the stars show a regular radial velocity 
curve, the gas radial velocity curve is asymmetric with respect to the center of the galaxy. There 
is a steep rise of $\mathit{V_{gas}}$ up to $220 \, \mathrm{km \, s^{-1}}$ within 1 kpc 
in the NW direction, whereas on the other side the gas velocity reaches only $160 \, \mathrm{km \, s^{-1}}$. 
We discover that the gas rotates also along the minor axis (see Fig.~\ref{fig:kin_A}).

\begin{figure}
 \centering
\resizebox{\hsize}{!}{\includegraphics{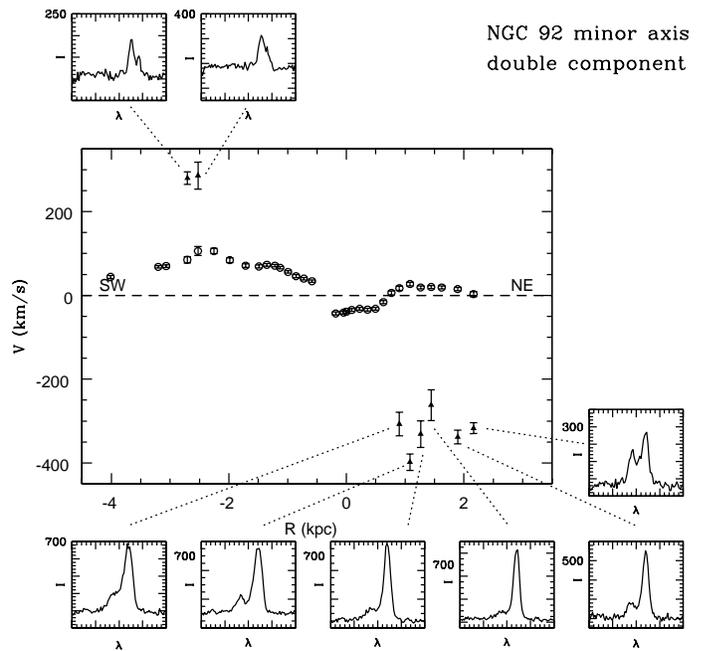}}
\caption{Position-velocity diagram with both the main (open circles) and the second (filled triangles)
component of gas along the minor axis of NGC 92. Smaller panels show the zoom of 1D spectra in the 
range 5040-5080 \AA\ containing the [OIII] emission-line at each position where we were able to 
measure both components.}
 \label{fig:doublecomp_A}
\end{figure}

Moreover, the emission line profiles of H$\beta$ and [OIII] $\lambda\lambda 4959, 5007$ along the 
minor axis reveal the presence of a secondary component in addition to the main one. 
In Fig.~\ref{fig:doublecomp_A} we show the position-velocity diagram with both the main 
(open circles) and the second (filled triangles) component of gas along the minor axis of NGC 92. 
Smaller panels show the zoom of 1D spectra in the range 5040-5080 \AA\ of the [OIII] emission-line at 
each position where we were able to measure both components. Near the nucleus the second component 
was difficult to separate due to the broadening of the line and it appeared as an asymmetry of the 
profile. As we go far away from the nucleus a second peak is clearly evident, as shown in the smaller 
panels of Fig.~\ref{fig:doublecomp_A}. At distances greater than $2.5 \, \mathrm{kpc}$ we 
do not detect the second component anymore.

\begin{figure}
 \centering
 \resizebox{\hsize}{!}{\includegraphics{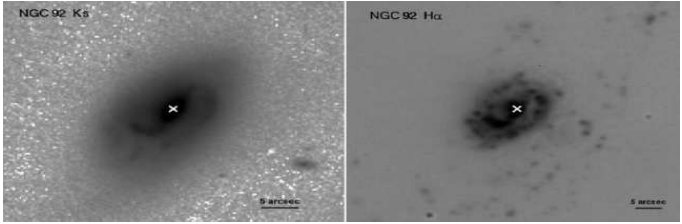}}
 \caption{Ks-band (left) and continuum-subtracted H$\alpha$+[NII] (right) images of NGC 92. 
 North is up and east to the left. The bar is well visible in the $Ks$ image. 
 Despite the different pixel scales of these images, it is clear that the H$\alpha$ bright knot 
 to the SE is the elbow of a spiral arm pointing towards the observer.}
 \label{fig:barra}
\end{figure}

The velocity gap between the components is $\Delta \mathit{V} \approx 200 \, \mathrm{km \, s^{-1}}$ 
in the SW direction whereas it reaches $\Delta \mathit{V} \approx 350 \, \mathrm{km s^{-1}}$ in the 
NE direction. There can be two possible explanations for this double component: either we are looking 
at infalling gas with extra-planar motions \citep{2008MNRAS.386..935F} or the two components trace 
the well known eight-shape due to the presence of a bar \citep{1995ApJ...443L..13K}. 
To understand which of the two proposed explanations is more likely, we
inspected the $Ks$-band and the continuum-subtracted H$\alpha$+[NII] images\footnote{The [NII]/H$\alpha$ emission line ratios evaluated in several regions of the galaxy by means of
lower resolution spectra (not presented in this paper) together with the shape of the narrow-band filter transmission curve show that the [NII] emission lines contribute roughly 11.5\% of the flux in the narrow-band image.} of NGC 92 (Hereafter H$\alpha$ refers to H$\alpha$+[NII]). The $Ks$-band image was obtained with SOFI at NTT and has a scale of 
$0.288 \arcsec \, \mathrm{px^{-1}}$, whereas the H$\alpha$ image was obtained with FORS2 and
has a scale of $0.1 \arcsec \, \mathrm{px^{-1}}$. These data will be presented and analyzed in detail 
in our forthcoming paper (Temporin et al. in preparation). Here we just report about a few
morphological details of the inner regions of NGC 92 that are particularly useful for our
interpretation of the kinematic features. The two images are shown in Fig.~\ref{fig:barra}.
The $Ks$ image clearly shows the presence of a bar, thus 
corroborating the hypothesis that the second gas component along the minor axis is due to 
a bar. Also, from the comparison with the H$\alpha$ image, it appears clearly that the H$\alpha$ 
bright knot to the SE of the galaxy nucleus, thought to be a secondary nucleus, 
is actually the elbow of an inner proceeding spiral arm.

In Sect.~\ref{sec:Discussion} we will explore and discuss the implications of the presence 
of a bar for NGC 92.

\begin{figure}
 \centering
\resizebox{\hsize}{!}{\includegraphics[angle=-90]{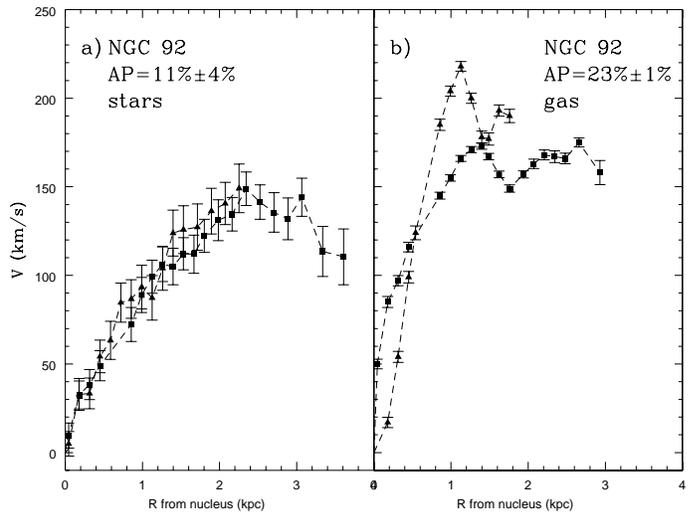}}
 \caption{Folding of the stellar (panel a) and the gas (panel b) radial velocity curve of NGC 92. 
 Triangles are used for the receding side, in NW direction and squares for the proceeding side, 
 in SE direction. The \textit{AP} values obtained for the stars and the gas are indicated.}
 \label{fig:AP_A}
\end{figure}

In Fig.~\ref{fig:AP_A} we show the folding of the stellar (panel a) and the gas (panel b) radial 
velocity curves and their \textit{AP}. The behavior of the stars looks quite regular, 
whereas the gas is much more perturbed. We obtain \textit{AP} values of 11\%$\pm$4\% and 
23\%$\pm$1\% for the stars and the gas, respectively. In a recent work \citet{2008A&A...484..299P} 
used simulations to analyze the \textit{AP} of interacting and unperturbed galaxies in CGs 
and compare it with observations. They found \textit{AP} $\approx$ 13-36\% for the formers and 
\textit{AP} $\approx$ 5-7\% for the latter. 
These two ranges of values are in good agreement with our estimates for the
observed velocity curves of the stars and the gas, respectively. Therefore, the stars of 
NGC 92 are unperturbed or only slightly perturbed while 
the gas \textit{AP} is in agreement with values obtained for other perturbed galaxies in groups. A mismatch between the two sides of the rotation curve was 
independently found also in a recent work by \citet{2009arXiv0908.2798T} 
based on Fabry-Perot data. They also found signs of secondary 
kinematical components
and a possible signature of non-circular motions.
We refer to Sect.~\ref{sec:Discussion} for a detailed discussion.

Additional emission lines are detected in the nucleus of NGC 92 and in the surrounding regions.
These are HeI $\lambda\lambda$ 5015.68, 5875.6, [NI] $\lambda\lambda 5198.5, 5200.7$ and 
[NII] $\lambda 5754.6$. Some of these lines require a sufficiently hard ionizing radiation
\citep{2006agna.book.....O} and, therefore, are normally found in regions photoionized by 
an AGN radiation, 
or in regions of strong and recent star formation, although most of them are generally 
difficult to detect because of their weakness. It is interesting to note that 
the (usually very weak) HeI $\lambda$ 5015.68 emission-line was detected only in correspondence 
of the elbow of the inner spiral arm, where additional data (not presented here) give evidence of 
the explosion of a supernova in year 1999 (Temporin et al., in preparation). 
The [NII] $\lambda 5754.6$ line -- which is a well known electronic temperature
indicator for the low ionization zones of gaseous nebulae, when combined with the more 
easily detectable [NII] $\lambda\lambda$ 6548,6583 
emission-lines -- is visible in the same spectrum and in the nuclear one. 
Finally, the [NI] doublet is detected in the nucleus and in all spectra out
to 1.4 kpc NW and 2.1 kpc SE of it along the major axis. 
Vaona et al. (in preparation), in a statistical study of the NLR of Sy2 galaxies
extracted from the SDSS, find that the [NI] $\lambda\lambda 5198.5, 5200.7$ doublet
is detected in 8\% of their sample (169 out of 2153 galaxies) and has a median ratio to 
H$\beta$ of $\sim$ 0.14 $\pm$ 0.06. This value is consistent with what we observe in the
nuclear region of NGC 92, which is classified as a LINER. Further out, the [NI]/H$\beta$ ratio
assumes values of the order of a few 10$^{-2}$, similar to those observed for HII regions
in other LINER galaxies \citep[e.g.][]{2000MNRAS.318..462D}. These [NI] lines are even 
observed in the presence
of relatively weak radiation fields, like in the partially ionized zones 
of the Orion Nebula, and fluorescence is thought to play an important role in their 
excitation \citep{1999ApJ...527..474B}.

\subsection{Kinematics of NGC 89}
\label{sec:kin89}

\begin{figure}
 \centering
 \resizebox{\hsize}{!}{\includegraphics[angle=-90]{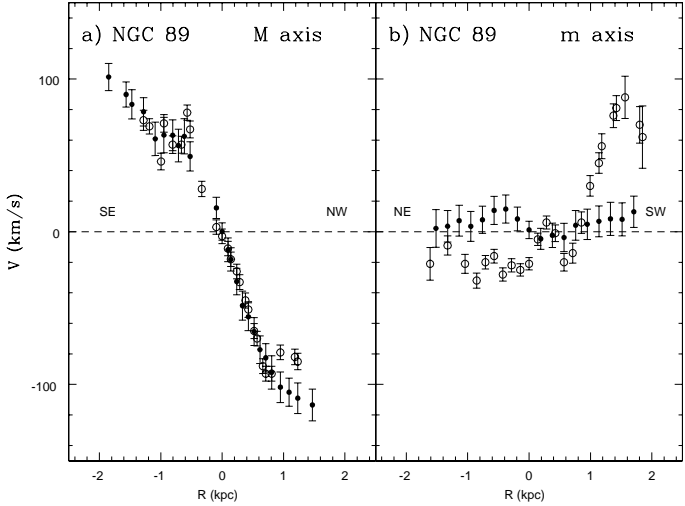}}
 \caption{Gas (open circles) and stellar (filled circles) kinematics along the Major (panel a) and 
 the minor (panel b) axis of NGC 89.}
 \label{fig:kin_B}
\end{figure}

In Fig.~\ref{fig:kin_B} we present the stellar (filled circles) and gas (open circles) 
radial velocity curves of NGC 89 along both the major (panel a) and the minor (panel b) axis. 
Along the major axis we can measure velocities only out to $2 \, \mathrm{kpc}$ from the nucleus 
and both the gas and the stars reach a maximum $\mathit{V} \approx 100 \, \mathrm{km \, s^{-1}}$. 
In the inner regions, $\mathit{R} < 1 \, \mathrm{kpc}$, gas and stars show similar kinematics, while further out the gas has a slightly less regular pattern. In particular we notice that major 
discrepancies 
occur at $\mathit{R} \approx 1 \, \mathrm{kpc}$ in the NW direction. The gas kinematics is perturbed 
also along the minor axis: in the SW direction the gas velocity starts to increase at 
$\approx 1 \, \mathrm{kpc}$ and reaches $\mathit{V} \approx 90 \, \mathrm{km \, s^{-1}}$ at 
$1.5 \, \mathrm{kpc}$. We suggest the possibility that the gas kinematics is affected 
by the central AGN and by the presence 
of the H$\alpha$-emitting extraplanar gas (see Sect.~\ref{sec:SCG0018-4854}).

We observe that the emission-line profile of [OIII] is asymmetric and irregular due to the presence 
of multiple components, at least three, in the nuclear region. This kind of profile is typically 
observed in the narrow line regions (NLR) of Sy2 galaxies. It is a common belief that photoionization 
and shocks 
play a dominant role in modifying the physical conditions of the gas
\citep{2002ApJ...572..753D,2000ApJ...544..763K,2005ARA&A..43..769V}, however the dynamics of the gas 
in the NLR is less certain. The presence of blue-shifts and red-shifts on either side of the nucleus 
indicates that rotation alone cannot explain the observed radial velocities. Emission-line profiles 
can be reproduced by different kinematic models: infall, outflow, rotation, etc. 
\citep{1986ARA&A..24..171O,1991ApJS...75..383V}. The general idea is that radial outflows play an 
important role.

\begin{figure}
 \centering
  \resizebox{\hsize}{!}{\includegraphics{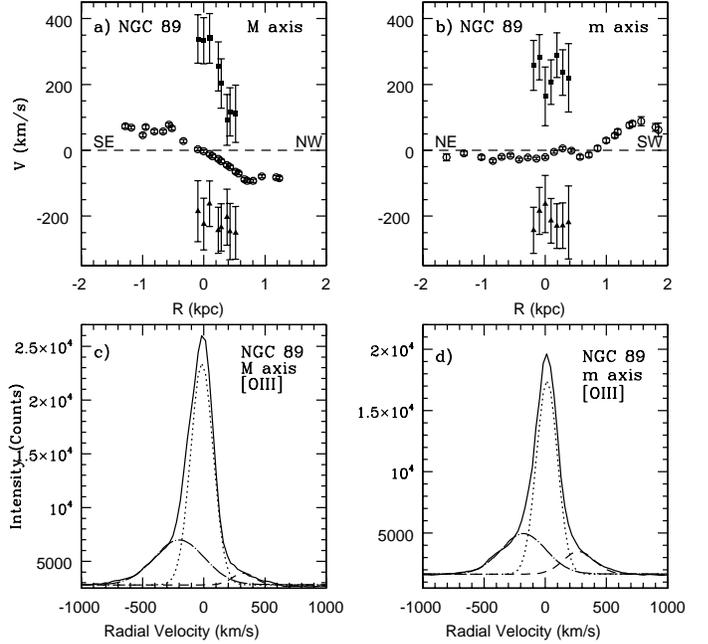}}
 \caption{NGC 89: position-velocity diagram along the major (panel a) and the minor (panel b) axis and 
 the [OIII] emission-line profile obtained by summing 6 pixels ($1\arcsec.2$) around the nucleus 
 along the major (panel c) and the minor (panel d) axis as a function of velocity with respect to galaxy 
 systemic velocity. We fit the profile with a combination of three Gaussians: the main component 
 (open circles and dotted line), the blue-shifted component (filled triangles and dot-dashed line) 
 and the red-shifted one (filled squares and dashed line).}
 \label{fig:multiplecomp_B}
\end{figure}

\begin{figure}
 \centering
 \resizebox{\hsize}{!}{\includegraphics[angle=-90]{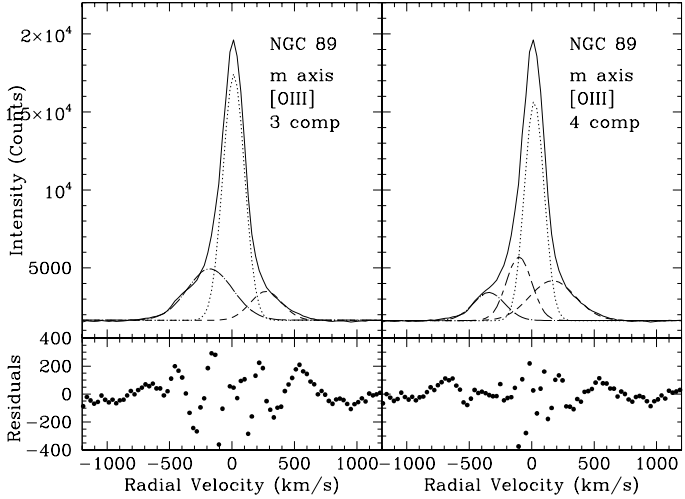}}
 \caption{Multiple component of the [OIII] emission-line profile along the minor axis of NGC 89. 
 We compare the fit of the profile using a combination of three (left) or four (right) Gaussians. 
 In the bottom panels we show residuals for both cases. The choice of four Gaussians, the main component 
 (dotted line), two blue-shifted components (dot-dashed and short-long dashed line) and the red-shifted one 
 (dashed line) gives better results.}
 \label{fig:fourcomp_B}
\end{figure}

The top panels of Fig.~\ref{fig:multiplecomp_B} show the position-velocity diagrams along 
the major (panel a) and the minor (panel b) axis for the main gaseous component (open circles), 
the [OIII] blue-shifted component (filled triangles), and the [OIII] red-shifted one (filled squares). 
We observe non-circular motions of the gas with high velocities, 
$\mathit{V} \approx \pm 200 \, \mathrm{km \, s^{-1}}$. 
In particular the red-shifted component along the major axis has a regular pattern with high 
velocities within $150 \, \mathrm{pc}$ from the nucleus and then decelerates as we go farther away, 
$500 \, \mathrm{pc}$, in analogy to other cases of Sy2 NLR kinematics 
\citep[see][and references therein]{2006AJ....132..620D}. The multiple components can be seen more 
clearly in the spectral extractions around the nucleus. The bottom panels of 
Fig.~\ref{fig:multiplecomp_B} show the [OIII] 
emission-line profile obtained by summing 6 pixels ($1\arcsec.2$) around the nucleus along the major 
(panel c) and the minor (panel d) axis as a function of velocity with respect to the systemic velocity
of the galaxy. We fit the profile with a combination of three Gaussians: the main component (dotted line), 
the blue-shifted component (dot-dashed line) and the red-shifted one (dashed line). The blue-shifted 
component along the minor axis fails to fit two different bumps in the [OIII] emission-line profile 
and a fourth component is probably needed to better reproduce the [OIII] profile. 
We are able to perform the fit with four components in a satisfactory way only in the center. 
In Fig.~\ref{fig:fourcomp_B} we show the comparison between the fit using three Gaussians (left) 
and the fit with four Gaussians (right). Residuals of the fit are shown in the bottom panel in both 
cases. The choice of four Gaussians, the main component (dotted line), two blue-shifted components 
(dot-dashed and short-long dashed line) and the red-shifted one (dashed line) gives better results, 
$\chi^{2}\approx1.7$, with respect to that of three components, $\chi^{2}\approx2.3$. 
The former residuals show a better agreement with the rms of the continuum as measured in two regions 
of 40 pixels at the edges of the emission-line profile. Although increasing the number of free 
parameters leads to better $\chi^{2}$, we should keep in mind that it also leads to a degeneracy of 
solutions. It is then difficult to establish which one is the most reliable, 3D spectroscopy with deeper integration time could help 
us to assess the possible presence of a fourth component.

However, it is interesting to note that this fourth component survives few pc away from the nucleus only 
in NE direction, that is the same direction of the galactic-scale H$\alpha$ brightest filament. 
We cannot exclude
that these two features are somehow linked to each other. The slit along the minor axis intercepts the 
jet-like filament in three different points. Therefore we are able to probe it 
spectroscopically and compare its kinematics with that of the adjacent gas. 

\begin{figure}
 \centering
 \resizebox{\hsize}{!}{\includegraphics{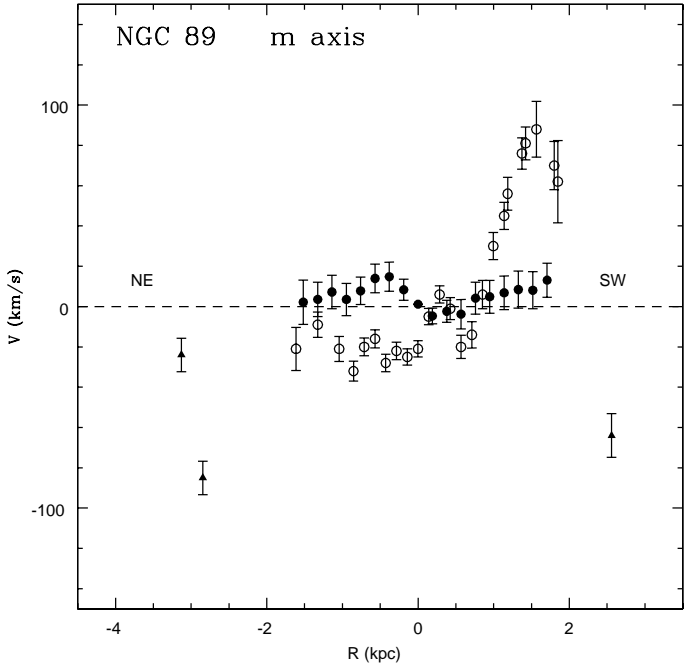}}
 \caption{Kinematics along the minor axis of NGC 89. Measured radial velocities for
 the primary gas component (open circles), the secondary gas component (filled triangles), 
and the stars (filled circles) are plotted. The secondary gas component is consistent
with the presence of an outflow.}
 \label{fig:outflow_B}
\end{figure}

In Fig.~\ref{fig:outflow_B} we plot the measured kinematics of the additional
gas components (filled triangles) together with the gas (open circles) and stars (filled circles) 
radial velocity curves. 
The kinematics of the extraplanar gas is clearly distinct from the disk gas kinematics.
In fact, in the SW direction this gas component shows opposite 
velocity with respect to the disk gas. In the opposite direction we have two measures 
and there is a possible indication that the secondary gas component slows down as we go far 
away from the center of the galaxy.
The decoupled kinematics of the secondary gas component is compatible with the
presence of an outflow. We discuss this possible interpretation in Sect.~\ref{sec:N89outflow}.

\begin{figure}
 \centering
 \resizebox{\hsize}{!}{\includegraphics[angle=-90]{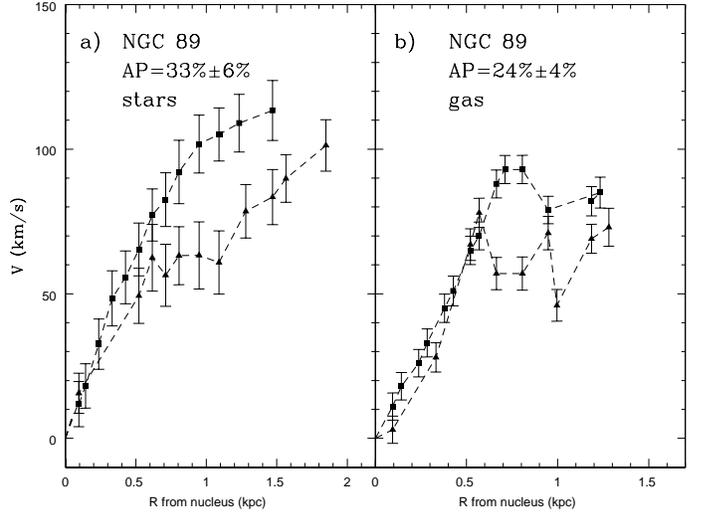}}
 \caption{Folding of the stellar (panel a) and gas (panel b) radial velocity curve of NGC 89. 
 Triangles are used for the receding side, in the SE direction and squares for the approaching side, 
 in the NW direction. The values of the \textit{AP} for the gas and the stars are indicated.}
 \label{fig:AP_B}
\end{figure}

Once again we have estimated the asymmetry parameter. In Fig.~\ref{fig:AP_B} we show the 
folding of the stellar (panel a) and the gas (panel b) radial velocity curve and the relative \textit{AP}. 
We obtain an \textit{AP} value of 33\%$\pm$6\% and 24\%$\pm$4\% for the stars and the gas respectively. 
These large values, as already mentioned in Sect.~\ref{sec:kin92}, are typical of strongly 
perturbed galaxies. We refer to Sect.~\ref{sec:Discussion} for a detailed discussion.

This galaxy as well shows many emission lines in the nuclear spectrum, in particular we reveal 
the presence of HeII $\lambda 4685.75$. 
This emission line can be observed in gas photoionized by the hard radiation field of an 
active galactic nucleus or it can be produced in the wind of Wolf-Rayet (W-R) stars 
\citep[e.g.][]{2007ARA&A..45..177C}, thus indicating a recent burst of massive star formation
\citep{1998ApJ...497..618S}. 
Usually, in W-R galaxies this emission-line
is broad (although accompanied by an often difficult to disentangle narrow nebular component) and 
forms the so-called ``W-R bump'' in the spectrum together with other 
strong emission features (e.g. of CIII, CIV, and NIII) in the wavelength range $\lambda$4650-4690 
\citep[][and references therein]{1998ApJ...497..618S}. Since the emission-line we observe is narrow, 
is not accompanied by a W-R bump and other typical features of W-R galaxies, and is visible only 
in the nuclear spectrum of NGC 89, we conclude that it is unlikely to be a Wolf-Rayet signature, but is 
rather the nebular emission produced by the ionizing radiation of the central AGN.

\subsection{Kinematics of NGC 87 and NGC 88}

In this Section we report the kinematical properties of the other two group members: 
NGC 87 and NGC 88. 

\begin{figure}
 \centering
 \resizebox{\hsize}{!}{\includegraphics[angle=-90]{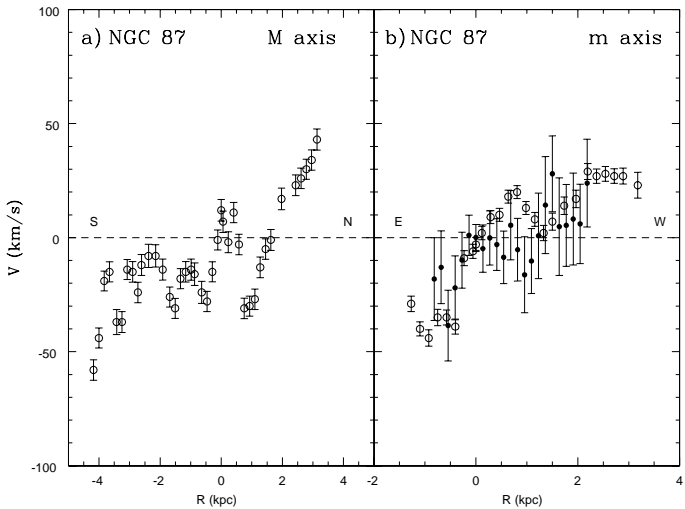}}
 \caption{Panel a -- Gas (open circles) kinematics along the Major axis of NGC 87. 
 Panel b -- Gas (open circles) and stellar (filled circles) kinematics along the minor axis of NGC 87.}
 \label{fig:kin_C}
\end{figure}

In Fig.~\ref{fig:kin_C} we present the stellar (filled circles) and the gas (open circles) 
radial velocity curves of NGC 87 along the minor axis (panel b), while along the major axis (panel a) 
we show only that of the gas (open circles). The lack of stellar kinematics along the 
major axis is due to problems in sky subtraction that made the use of XCSAO impossible, 
see Sect.~\ref{sec:method}. The gas shows a velocity gradient along both axes above which 
peculiar motions are superimposed. These peculiar motions belong to different regions of emission 
of the irregular galaxy. In both cases the velocity gradient is a regular trend with a significant 
velocity variation with respect to the errorbars and the velocity range of peculiar 
motions. Nevertheless, we cannot disentangle if the velocity gradient is associated to the whole galaxy 
or a few regions or if it is the evidence of a warp. The HI gas emission is more extended than the 
optical light and the emission looks asymmetric; the HI velocity field does not show any significant 
gradient, as it would be for an almost face-on galaxy, except in the N/NE corner at approximately 
the same position where the velocity gradient measured in the optical shows the largest variation.

\begin{figure}
 \centering
 \resizebox{\hsize}{!}{\includegraphics{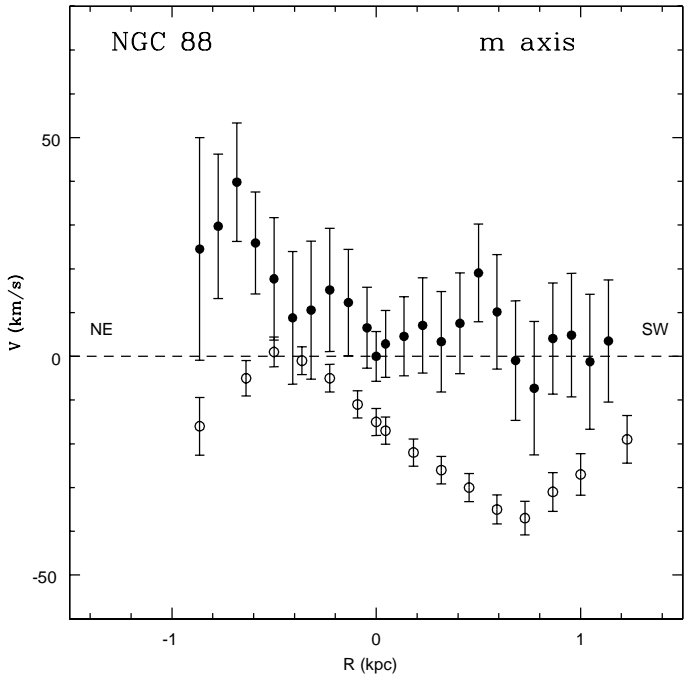}}
 \caption{Gas (open circles) and stellar (filled circles) kinematics along the minor axis of NGC 88.}
 \label{fig:kin_D}
\end{figure}

In Fig.~\ref{fig:kin_D} we show the stars (filled circles) and gas (open circles) kinematics 
along the minor axis of NGC 88. The gas shows a sinusoidal pattern possibly indicating a regular rotation. 
Once again, the ionized gas is moving on extra-planar orbits. This kind of kinematics is widespread in 
galaxy bulges \citep{2004A&A...416..507C} and it could be generated by a triaxial bulge or a bar. 
Kinematical information along the major axis is needed to find out which hypothesis is more
reliable \citep{2003A&A...408..873C}.

\section{Discussion}
\label{sec:Discussion}

\subsection{Kinematical effects of the interaction on involved galaxies}

NGC 92 has clearly interacted with one or several group members as suggested by the extended tidal tail. 
We discuss now the kinematical effects of this interaction and which could be the other interacting galaxy/ies. 

NGC 92 stellar radial velocity curve is only moderately perturbed, whereas the gas has an asymmetric radial 
velocity curve. The receding side has a steeper rise than the approaching one. Note that the 
 approaching side is in the direction of the tidal tail, we suppose that the tail affects the gas 
kinematics keeping velocities at an approximately costant value of $160 \, \mathrm{km \, s^{-1}}$. 
This is supported by HI data: neutral hydrogen velocities show a continuous trend between the tidal tail 
and the approaching side of the galaxy disk, clearly suggesting a physical link between 
gas, disk, and tail \citep{2007A&A...473..399P}. Arp 295 is a similar case of interacting galaxies with 
asymmetric rotation curves \citep{2007RMxAA..43..179R}. Also for Arp 295 the rotation curve rises slowly 
and monotonically on the side pointing towards the tail and the companion whereas on the other side it 
rises more steeply. Furthermore, a study of a sample of interacting galaxies reveals that 
there is a correlation 
between the SFR and perturbed kinematics \citep{1993AJ....106.1771K}. The SFR increases as the rotation 
curves become more asymmetric from normal to disturbed rotation curves. 
Therefore, we expect to find enhanced 
star formation for NGC 92. Previous results from an H$\alpha$ study of SCG0018-4854 reveal that the 
luminosity function of the HII regions of NGC 92 exhibits a considerable flattening at the high 
luminosity end \citep{2005ASSL..329P..78T}. NGC 89 and NGC 88 are candidate interacting galaxies with 
NGC 92, in fact we find irregular radial velocity curves and peculiar kinematics for both galaxies.

NGC 89 has perturbed stellar and gas radial velocity curves suggesting that it has interacted with other
group members. 
Both gas and stars show lower velocities for the receding side with respect to the approaching one. 
This could be the result of different phenomena: (1) a disk warp, (2) streaming motions that are induced by 
interaction (3) composition of disk and spiral arm motions. 3D spectroscopy, which is currently 
unavailable for this galaxy, would help disentangling among these possibilities. Anyway, it is interesting 
to note that neutral and molecular hydrogen studies of this galaxy support the hypothesis that NGC 89 has 
interacted in the past with other group members. In fact NGC 89 has been completely stripped of its 
HI gas \citep{2007A&A...473..399P} and CO observations show that it is perturbed also in its molecular gas
distribution \citep{1996A&A...314..738B}.
These facts suggest that also the perturbed kinematics of NGC 89 is a consequence of past
interactions, the most likely perturber being NGC 92.

NGC 88 shows a sinusoidal pattern of the gas velocities probably due to a triaxial bulge or a nuclear bar. 
Both phenomena could be the result of an interaction. Moreover, there is an HI bridge between NGC 92 
and NGC 88 \citep{2007A&A...473..399P}. Broad band observations show a wide stellar tail departing from 
this galaxy towards south-west (Temporin et al. in preparation). 

The interaction process probably involves NGC 87 only partially: we found velocity gradients along both axis 
which could be the result of a slight interaction with other members. A stronger encounter would probably 
disrupt this galaxy due to its irregular morphology and lower mass.

\subsection{The interaction strength and history of SCG0018-4854 galaxies}

The suggestion of a double nucleus for NGC 92 \citep{1981A&A....98..223D} could give rise to the hypothesis 
that this galaxy is a merger remnant. If this were the case, NGC 92 should have interacted violently
with a companion galaxy in the past. Our results show that even if the interaction has been strong, 
we can exclude the hypothesis of a merger for NGC 92. The discovery of the nuclear bar rejected the 
hypothesis of a double nucleus. The comparison between $Ks$-band and H$\alpha$ images shows that the 
bright knot is actually the elbow of the inner proceeding spiral arm that departs from the nuclear bar. 
Therefore, NGC 92 should have interacted more recently and less strongly than the presence of a double nucleus could suggest.

However, it is well known that structures such as bars and rings could arise from galaxy 
interactions \citep{1961SvA.....5..376L,1987MNRAS.228..635N,1990A&A...230...37G}.
A direct tidal interaction acts as a destabilizing agent reducing the bar formation time-scale 
and accelerating the growth of the bar. Hence, we cannot exclude the possibility that 
the nuclear bar of NGC 92 be a product of its interaction with other group 
members and the presence of the tidal tail strengthens this hypothesis. 
3D-spectroscopy could help to better understand 
the origin of the bar: e.g. an evidence of gas inflow could be a proof of a recent bar formation. 
Furthermore, a detailed study of the star formation along the bar could help to state whether 
the bar is linked to the interaction or not.

We try to quantify the interaction strength estimating the asymmetry parameter
for NGC 92 and NGC 89 radial velocity curves. Cluster galaxies have a mean \textit{AP} of
12.6\%$\pm$1.2\% \citep{2001AJ....121.1886D} while, as we already said, galaxies
in groups have larger values. We find \textit{AP} values in the range 11-33\% which are
in agreement with values of other perturbed and interacting galaxies in CGs. The gas
is the first component that is affected in an interaction. Actually both NGC 92
and NGC 89 have been strongly perturbed in their gas kinematics. On the contrary,
stellar kinematics are perturbed if the interaction has been violent. Therefore,
since only NGC 89 shows a perturbed stellar kinematics, we speculate that the interaction
was stronger for this galaxy than for NGC 92. Furthermore taking into account that NGC 89 is the only galaxy of the group depleted of HI we
can not exclude a more complicated interaction history for this galaxy. NGC 89 could have interacted with the other groups members several
times being  NGC 92 the latest one. Otherwise NGC 89 could have joined the group only recently: the interaction with the
group potential removed its gas and then it interacted with NGC 92. Hydrodynamical simulations could help us drawing the more likely
scenario. Some caution in the interpretation is necessary, as recent studies show that
the appearance of distortions in rotation curves is strongly dependent on the
viewing angle \citep{2006A&A...458...69K}. In any case there is a clear trend of
increasing \textit{AP} for interacting galaxies. \citet{2008A&A...484..299P} 
analyzed the evolution of \textit{AP} in simulations of galaxy
encounters with varying set-up. They showed that \textit{AP} rises at the 
peri-passage with values up to $\approx$
20\% and it starts decreasing only  after $0.7 \, \mathrm{Gyr \, \mathit{h^{-1}_{70}}}$. In other words \textit{AP}
could be used as a timer for recent interactions. 
Being SCG0018-4854 a compact group, its galaxies are supposed to encounter many times 
during their lives. We try to determine the age of the latest interaction among the galaxies 
of this group using the \textit{AP}. Our estimated values of \textit{AP} for the
gas and stars of NGC 92 and NGC 89 are large and suggestive of a recent
interaction. Hence NGC 92 and NGC 89 should have interacted within 
the last $0.7 \, \mathrm{Gyr}$. This is only an upper limit as dynamical considerations on the 
tidal tail lead to a more recent interaction. A rough estimate of the age of the latest 
interaction is given by $\mathit{\tau_{tail}}\:\approx\:\mathit{L_{tail}/V_{flat}}$, where 
$\mathit{L_{tail}}$ is the tail extent, $\approx 30 \, \mathrm{kpc}$, and $\mathit{V_{flat}}$ 
is the velocity at which rotation curves become flat. In the tail direction V assumes an approximately 
constant value of $\approx 160 \, \mathrm{km \, s^{-1}}$, which gives 
$\mathit{\tau_{tail}} \approx 190 \, \mathrm{Myr}$. 

The above considerations lead to an estimated age of the latest interaction, 
$\approx 0.2 \, \mathrm{Gyr} < \mathit{\tau_{coll}} < 0.7 \, \mathrm{Gyr}$, 
which is comparable with the crossing time of this group: $0.272 \, \mathrm{Gyr}$ \citep{2007A&A...473..399P}. 
Further constraints on this matter will be obtained from the estimate of the age of the stellar populations through 
a detailed photometric analysis that is the subject of the already mentioned forthcoming paper.

\subsection{A galactic-scale outflow in NGC 89}\label{sec:N89outflow}

As mentioned in Sect.~\ref{sec:SCG0018-4854}, the continuum-subtracted H$\alpha$ image 
of NGC 89 shows the presence of ionised extra-planar gas \citep{2005ASSL..329P..78T}. 
This gas appears to be organised in a
biconical structure departing from the central part of the disk and 
extending roughly perpendicularly to the disk plane on both sides of the disk.
It has a filamentary structure and the prominent 4-kpc long jet-like structure
is possibly just the brightest of these filaments.    
The origin and the nature of these H$\alpha$ filaments 
cannot be fully established with the data presently available,  
but their geometry suggests the hypothesis of a galactic-scale outflow.
Such an outflow could be driven by the combined action of the AGN and
a central starburst, whose presence is indicated by a ring of bright 
H$\alpha$ emitting regions in the central part of the galaxy. 
Indeed, the X-ray emission from NGC 89 was modelled as an absorbed AGN plus a diffuse 
emission coming from the central starburst \citep{2008A&A...484..195T}. 
However, we cannot exclude the possibility of an inflow. In order to discriminate
between an inflow and an outflow beyond any doubt, we would need a detailed 
study of the interstellar absorption-line profiles, such as the NaI D $\lambda\lambda$
5890, 5896 \AA\, \citep[e.g.][]{2005ApJ...632..751R}.

Although another possibility would be the hypothesis that a minor merger 
has occurred in a recent past and a dwarf satellite has been cannibalised by NGC 89, we find
this interpretation  less likely. In fact, at least with the existing data, we could
not find any trace of a secondary nucleus in NGC 89. A search of the 
NED archive for the presence of dwarf satellites around the group revealed
only the presence of two galaxies within an Abell radius with comparable redshift.
This suggests that SCG0018-4854 is genuinely isolated. Still, we cannot
exclude that NGC 89 might have had a dwarf satellite that was disrupted in 
a mergeer event. However, in the case of a
minor merger we would expect to observe structures such has tidal streams
forming loops and shells around the main galaxy or features resembling X-shaped
jets, as it is the case for other systems that have been explained as the result
of the disruption of a dwarf galaxy cannibalized by its more massive companion
\citep[e.g. Mrk 315 and NGC 1097, see][]{2005MNRAS.360..253C,2003ApJ...585..281H}.
Instead, the H$\alpha$ images show a wide distribution of extra-planar gas on
both sides of the galaxy disk.

On the other hand, both AGN-driven and starburst-driven winds on galaxy scale
have been observed in other Seyfert 2 and starburst galaxies both in the
local Universe and at higher redshifts 
\citep[see][for a review]{2005ARA&A..43..769V}.
Powerful large-scale outflows might be caused by
strong starbust activity in the central regions of a galaxy with a consequent
shock-heating and acceleration of ambient interstellar gas \citep[e.g.][]{1990ApJS...74..833H}. 
Recent simulations 
of starburst-driven galactic winds reveal that optically emitting filaments can be formed by 
fragmentation and acceleration of a cloud into a supersonic wind \citep{2008ApJ...674..157C}. 
When radiative cooling is taken into account, these fragments survive enough to generate strands and 
filaments downstream of the original cloud position like those seen in starburst-driven 
winds \citep{2009arXiv0907.4004C}.
A well known case is that of M82,
which shows optical filaments whose velocity increases with radius \citep{1998ApJ...493..129S}. 
The observed morphologies
and kinematics of outflowing gas on large scales reminds of the structures
we observed in NGC 89. By analysing a sample of infrared
luminous starbursts, \citet{2005ApJS..160..115R} found that superwinds are 
ubiquitous in these galaxies and in most
cases the outflowing gas has velocities of 100 - 200 km $s^{-1}$. 

\citet{1997AJ....113.1678B,2001A&A...380...40T,2004ApJS..151..193S} 
have shown the presence of a correlation between H$\alpha$ outflow structures and X-ray features. 
\citet{2009arXiv0907.4004C} demonstrate also that the soft X-ray emission is due to mass ablated 
from the clouds and mixed into the surrounding gas. After $0.75 \, \mathrm{Myr}$ the fraction 
of gas mixed into the surrounding gas is $\approx 25$\%.
Hence, we would expect to observe some X-ray features correlated to the galactic-scale 
filaments of NGC 89. \citet{2008A&A...484..195T} did not detect any X-ray feature linked to the 
H$\alpha$ emission. 
However, the total number counts of their X-ray observations and the presence of 
two dominant components in the emission related to NGC 89, namely an absorbed AGN and 
an unresolved component from binary systems, would make it extremely difficult to detect 
a further faint component related to the low surface brightness H$\alpha$ filaments that 
trace the extra-planar gas ($\mathit{Flux_{H\alpha}} = 10^{-16}-10^{-17} \, \mathrm{erg \, 
s^{-1} \, cm^{-2}}$). Also, any interaction of the outflow with the IGM is expected
to be very weak, since the IGM in which the members of SCG0018-4854 are embedded 
is cooler ($\mathit{kT} \approx 0.2 \, \mathrm{keV}$) and less dense 
($\mathit{n_e} \approx 5 \times 10^{-3} \, \mathrm{cm^{-3}}$) than that typically found in 
other environments such as groups with a central elliptical galaxy or the cores 
of clusters \citep{2008A&A...484..195T}.

Our study of NGC 89 reveals a complex gas kinematics in the inner regions:
emission line profiles have multiple components along both the major and the minor axis.
We show evidences of at least two secondary components along both axis: one
blue-shifted and the other red-shifted. This kind of composite motions are well
known in literature for the NLR of Seyfert galaxies \citep[][and references therein]{2006AJ....132..620D}. 
However, a detailed kinematical model for
the NLR is beyond the scope of this paper.
The presence of a possible fourth component in the emission-line profile along 
the minor axis is intriguing as this component is visible only in the direction of the 
galactic-scale extra-planar features. If indeed a galactic-scale outflow is
present, we cannot exclude a link between such an outflow and the central engine.

In Sect~\ref{sec:kin89} we probed the kinematics of the extra-planar gas in
a few positions.
The velocity difference of $\approx 60 \, \mathrm{km \, s^{-1}}$ between the two measured
positions in NE direction could be interpreted as a signature of the presence of two 
gas shells with different velocities due to the impact of the gas with the ISM. 
Moreover, when going far away from the nucleus along the minor axis we found that 
the gas is accelerated in the SW direction, consistently with the outflow interpretation.

Therefore, the combined information stemming from the kinematics and
the geometry of the ionised extraplanar gas favours the outflow scenario.
\citet{1998A&A...333..459M} studied a similar case: NGC 2992 is a Seyfert 1.9 galaxy with 
an emission arc, visible in [OIII] and H$\alpha$, in interaction with its companion NGC 2993. 
A radial outflow was added to the circular motion to account for the observed kinematics. 
They found largest differences with respect to the model when going at larger distances
from the nucleus
and suggested that these are kinematically distinct regions where the gas could be perturbed 
by the interaction of NGC 2992 with its close companion NGC 2993. 
This could be the case also for NGC 89 since the HI velocity field reveals that the 
tidal tail of NGC 92 is pointing towards the observer with a velocity comparable to that 
of NGC 89 \citep{2007A&A...473..399P}

With the currently available data we are not able to establish the origin 
of the galactic-scale outflow. 
Integral-field spectroscopy would be useful to better understand the outflow kinematics
and constrain its origin.

\subsection{Comparing SCG0018-4854 with HCG16}

The extreme physical properties of SCG0018-4854 make it a very interesting and
rare group. It could be considered the southern counterpart of HCG16, the group
with the highest concentration of late-type active and interacting galaxies in
the nearby universe. Therefore, we compare its kinematic properties with
those found in literature for HCG 16. The latest and more detailed kinematical
study of HCG16 has been carried out by \citet{1998ApJ...507..691M} using
Fabry-Perot data. They found kinematic peculiarities for three galaxies out of
the four originally listed in Hickson's catalogue \citep{1982ApJ...255..382H}.
They found a non-axisymmetric rotation curve with a velocity discrepancy of $100
\, \mathrm{km \, s^{-1}}$ between the two sides and a slow rise in the central region. They
discovered the presence of a second component in the emission line profile of
one galaxy. They detected severe warping of the kinematic major axis of two
galaxies and finally the presence of a double nucleus and a peculiar bar-like
structure. Based on these results, they suggested that major merger events 
have taken place in at
least two out of four galaxies. Our study reveals that most of these kinematic
peculiarities are also present in SCG0018-4854, confirming its nature of a
relatively dynamically young and active CG.

\section{Conclusions}
\label{sec:Conclusions}

We have analyzed high signal-to-noise spectra along both the major and the minor axis 
of each galaxy of SCG0018-4854. Each galaxy shows disturbed or peculiar kinematics.
Kinematic information is important in order to have some insight about the possible interaction 
history of the galaxies. Different interaction scenarios, depending on the 
strength and the geometry of the encounter and the morphological types of the interacting 
systems, will leave different signatures on the galaxy kinematics. Following 
\citet{1998ApJ...507..691M} we found out signs of disturbed velocity fields, asymmetric 
rotation curves, multiple kinematic gas components, tidal tails and nuclear activity. 
All these characteristics suggest that at least three out of four galaxies of SCG0018-4854 
have been involved in a strong and recent interaction. We derived some constraints for 
the age of the latest close interaction: $\approx 190 \, \mathrm{Myr} < \mathit{\tau_{coll}} 
< 0.7 \, \mathrm{Gyr}$. The interaction has been strong enough to perturb the gas kinematics 
up to a level of 24\% for the gas of the two main galaxies. These results are in agreement with the recent claim by 
\citet{2009arXiv0908.2798T} that this group is in an advanced stage of 
interaction, based on the presence of young star forming regions and a 
candidate tidal dwarf galaxy
identified in the UV in addition to the observed tidal tails. Finally, the results of our spectroscpic
analysis are consistent with the hypothesis of a large-scale outflow in NGC 89.

In contrast to its evolved properties as a group such as the high spatial density, 
the spatial distribution of HI and the presence of a common hot IGM, SCG0018-4854 is
entirely composed of spiral galaxies and has 
a remarkably high fraction of active galaxies. Moreover, this study reveals 
kinematical signs of recent interactions among its members making SCG0018-4854 
a dynamically young group although its global properties suggest a more advanced stage of evolution.

Given the estimated age of the latest interaction, we could say that we catch these 
galaxies in the act of interacting. What about the future of this group? 
Hydrodynamical simulations could help us to understand the fate of SCG 0018-4854, 
a really isolated interacting compact group.

\begin{acknowledgements}
We are grateful to the anonymous referee for his/her prompt and 
constructive
comments that helped us to improve this paper.
This research has
made use of the NASA/IPAC Extragalactic Database (NED) which
is operated by the Jet Propulsion Laboratory, California Institute of
Technology, under contract with the National Aeronautics and Space 
Administration. V.P. is grateful for support from the short term studentship DGDF 09/39 awarded by the European Southern Observatory (ESO).
\end{acknowledgements}

\bibliographystyle{aa} 
\bibliography{Presotto_2009_scg0018}

\end{document}